**EMPIRICAL RESEARCH**  **Open Access**

# A neural network-supported two-stage algorithm for lightweight dereverberation on hearing devices

Jean-Marie Lemercier[1*], Joachim Thiemann[2], Raphael Koning[2] and Timo Gerkmann[1]

**Abstract**

A two-stage lightweight online dereverberation algorithm for hearing devices is presented in this paper. The approach combines a multi-channel multi-frame linear filter with a single-channel single-frame post-filter. Both components rely on power spectral density (PSD) estimates provided by deep neural networks (DNNs). By deriving new metrics analyzing the dereverberation performance in various time ranges, we confirm that directly optimizing for a criterion at the output of the multi-channel linear filtering stage results in a more efficient dereverberation as compared to placing the criterion at the output of the DNN to optimize the PSD estimation. More concretely, we show that training this stage end-to-end helps further remove the reverberation in the range accessible to the filter, thus increasing the *early-to-moderate* reverberation ratio. We argue and demonstrate that it can then be well combined with a post-filtering stage to efficiently suppress the residual late reverberation, thereby increasing the *early-to-final* reverberation ratio. This proposed two-stage procedure is shown to be both very effective in terms of dereverberation performance and computational demands, as compared to, e.g., recent state-of-the-art DNN approaches. Furthermore, the proposed two-stage system can be adapted to the needs of different types of hearing-device users by controlling the amount of reduction of early reflections.

## 1 Introduction

Communication and hearing devices require modules aiming at suppressing undesired parts of the signal to improve the speech quality and intelligibility. Reverberation is one of such distortions caused by room acoustics and is characterized by multiple reflections on the room enclosures. Late reflections particularly degrade the speech signal and may result in a reduced intelligibility [1].

Traditional approaches were proposed for dereverberation such as spectral enhancement [2], beamforming [3], a combination of both [4], coherence weighting [5, 6], and linear-prediction based approaches such as the well-known weighted prediction error (WPE) algorithm [7, 8]. WPE computes an auto-regressive multi-channel filter in the short-time spectrum and applies it to a delayed group of reverberant speech frames. This approach is able to partially cancel late reverberation while inherently preserving parts of the early reflections, thus improving speech intelligibility for normal and hearing-supported listeners [9].

WPE and its extensions require the prior estimation of the anechoic speech PSD, which is modeled for instance through the speech periodogram [7] or a power-compressed periodogram corresponding to sparse priors [8], by an autoregressive process [10] or through non-negative matrix factorization [11]. A DNN was first introduced in [12] to model the anechoic PSD, thus avoiding the use of an iterative refinement.

*Correspondence:
Jean-Marie Lemercier
jeanmarie.lemercier@uni-hamburg.de
[1] Signal Processing, Universität Hamburg, Hamburg, Germany
[2] Advanced Bionics, Hannover, Germany

Instead of providing parameters for linear prediction as in, e.g. [12, 13], DNNs were also proposed for mapping-based dereverberation in the time-frequency magnitude domain [14], complex domain [15, 16], or in the time-domain [17].

As hearing devices operate in real-world scenarios in real-time, the proposed techniques for dereverberation should support low-latency online processing and adapt to changing room acoustics. Such online adaptive approaches were introduced, based on either Kalman filtering [18, 19] or on a recursive least squares (RLS)-adapted WPE, which is a special case of Kalman filtering [20]. Strategies for handling the case of speakers changing positions were introduced in [19, 20]. In the RLS-WPE framework, the PSD is either estimated by recursive smoothing of the reverberant signal [20] or by a DNN [21].

In the previously cited works, the DNN is trained towards PSD estimation, although this stage is only a front-end followed by RLS-WPE-based dereverberation algorithms. So-called end-to-end techniques aim to solve this mismatch by using a criterion placed at the output of the complete algorithm to train the DNN. End-to-end techniques using an automatic speech recognition (ASR) criterion were designed to refine the front-end DNN handling, e.g., speech separation [22], denoising [23], or multiple tasks [24]. An end-to-end procedure using ASR as a training criterion was also introduced in [25] to optimize a DNN used for online dereverberation.

This journal paper is an extension of our prior work [26], where we proposed instead to use a criterion directly on the output signal rather than using ASR. We experimentally showed that it improved instrumentally predicted speech intelligibility and quality. The proposed criterion also enabled us to use different target signals and corresponding WPE parameters to make our approach adapt to the needs of different hearing-aid users categories: hearing aid (HA) users on the one hand benefiting from early reflections like normal listeners [9] and cochlear implant (CI) users on the other hand which do not benefit from early reflections [27].

We noticed in [26] that although the energy residing in the moderate reverberation range corresponding to the filter length was particularly suppressed when training the approach end-to-end, residual late reverberation could still be heard at the output. A further processing stage could be dedicated to removing this residual reverberation, as increasing the length of the linear filters results in rapidly increasing computational complexity. Hybrid approaches using such cascaded DNN-assisted stages have been proposed for dereverberation [28] or joint dereverberation, separation, and denoising [13, 24, 29].

The extension to our work [26] consists in the three following contributions. First, we introduce metrics to measure the energies in various reverberation ranges in order to investigate the differences between the previously cited WPE-based approaches and our proposed method. Second, we propose to use a second DNN-supported stage based on single-frame non-linear magnitude filtering and show that it significantly suppresses the residual late reverberation at the output of WPE. We show with the newly introduced metrics that this latter stage particularly benefits from strong dereverberation within the linear filter range obtained with the previous end-to-end WPE approach. Finally, we evaluate our approach and baselines on simulated reverberant data inspired by the WHAMR! dataset [30].

The rest of this paper is organized as follows. In Section 2, the online DNN-WPE dereverberation scheme is summarized. Section 3 presents the DNN-supported postfilter and describes the used end-to-end training procedure. In Section 4, we describe the experimental setup and introduce metrics in order to detail the dereverberation performance in various ranges. The results are presented and discussed in Section 5.

## 2 Signal model and DNN-supported WPE Dereverberation

### 2.1 Signal model

We use a subband-filtering approximation in the shorttime Fourier transform (STFT) domain as in [7], and all computations except those involving neural networks are computed for each frequency band independently. Therefore, we omit the frequency index $f$ when unnecessary and all vectors and matrices have an additional implicit frequency dimension of size $F$. The time frame index in the sequences of length $T$ is denoted by $t$ and is also dropped when not explicitly needed. We use lowercase normal font notation for signals having only time (and frequency) dimensions ($a_t \in \mathbb{C}$), lowercase bold font notation for vectors having one extra dimension ($\boldsymbol{a}_t \in \mathbb{C}^{d_1}$) and reserve uppercase bold font notation for matrices having two extra dimensions ($\boldsymbol{A}_t \in \mathbb{C}^{d_1 \times d_2}$).

The reverberant speech $\boldsymbol{x} \in \mathbb{C}^{D \times T}$ is obtained at the $D$-microphone array by convolution of the anechoic speech $s \in \mathbb{C}^T$ and the room impulse responses (RIRs) $\boldsymbol{h} \in \mathbb{C}^{D \times N}$:

$$\boldsymbol{x}_t = \sum_{\tau=0}^{N-1} \boldsymbol{h}_\tau s_{t-\tau} + \boldsymbol{u}_t = \boldsymbol{d}_t + \boldsymbol{e}_t + \boldsymbol{r}_t + \boldsymbol{u}_t, \quad (1)$$

where $\boldsymbol{d}$ denotes the direct path, $\boldsymbol{e}$ the early reflections component, $\boldsymbol{r}$ the late reverberation, and $\boldsymbol{u}$ an error term comprising modeling errors and background noise. The early reflections component $e$ was shown to contribute to speech quality and intelligibility for normal and HA listeners [9] but not for CI users, particularly in highly-reverberant scenarios [27]. Therefore, we propose that the dereverberation objective is to retrieve $v = d + e$ for HA listeners and $v = d$ for CI listeners.

## 2.2 WPE dereverberation

In relation to the subband reverberant model in (1), the WPE algorithm [7] uses an auto-regressive model to approximate the late reverberation $r$. Based on a zero-mean time-varying Gaussian model on the STFT anechoic speech $s$ with time (and frequency) dependent PSD $\lambda^{(\mathrm{WPE})}$, a multi-channel filter $G \in \mathbb{C}^{DK \times D}$ with $K$ taps is estimated. This filter aims at representing the inverse of the late tail of the RIRs $h$, such that the target $v$ can be obtained through linear prediction with delay $\Delta$. The prediction delay $\Delta$ is originally intended to avoid undesired short-time speech cancelations in [7]; however, this also leads to preserving parts of the early reflections. As such, we propose to set $\Delta$ larger for normal hearing and HA users who benefit from early reflections [9] but lower for CI users who suffer from early reflections [27]. By disregarding the error term $u$ in (1) in noiseless scenarios, we obtain:

$$v_t^{(\mathrm{WPE})} = x_t - G_t^H \mathcal{X}_{t-\Delta}, \qquad (2)$$

where $\mathcal{X}_{t-\Delta} = \left[ x_{t-\Delta}^T, \ldots, x_{t-\Delta-K+1}^T \right]^T \in \mathbb{C}^{DK}$.

In order to obtain an adaptive and real-time capable approach, RLS-WPE was proposed in [20], where the WPE filter $G$ is recursively updated along time. RLS-WPE can be seen as a special case of Kalman filtering, in which the target covariance matrix is replaced by the scaled identity matrix $\lambda^{(\mathrm{WPE})}I$, and the weight state error matrix is simply updated by dividing by the recursive factor $\alpha$ instead of following the usual Markov model [19]:

$$k_t = \frac{(1-\alpha) R_{t-1}^{-1} \mathcal{X}_{t-\Delta}}{\alpha \lambda_t^{(\mathrm{WPE})} + (1-\alpha) \mathcal{X}_{t-\Delta}^H R_{t-1}^{-1} \mathcal{X}_{t-\Delta}}, \qquad (3)$$

$$R_t^{-1} = \frac{1}{\alpha} R_{t-1}^{-1} - \frac{1}{\alpha} k_t \mathcal{X}_{t-\Delta}^T R_{t-1}^{-1}, \qquad (4)$$

$$G_t = G_{t-1} + k_t (x_t - G_{t-1}^H \mathcal{X}_{t-\Delta})^H. \qquad (5)$$

$k \in \mathbb{C}^{DK}$ is the Kalman gain, $R \in \mathbb{C}^{DK \times DK}$ the covariance of the delayed reverberant signal buffer $\mathcal{X}_{t-\Delta}$ weighted by the PSD estimate $\lambda^{(\mathrm{WPE})}$, and $\alpha$ the forgetting factor.

In non-idealistic scenarios, the term $u$ is not zero. Therefore, a regularization parameter $\epsilon > 0$ is added to the denominator of (3) which can be seen as a form of spectral flooring as used in traditional spectral enhancement schemes [4, 6, 31]. Although it is not per se a denoising solution and we still consider scenarios where noise is negligible in comparison to reverberation, adding this parameter helps increasing the robustness of WPE to noise, numerical instabilities and modeling errors. On the other hand, setting $\epsilon$ to a high value will excessively attenuate the relative variations of the Kalman denominator, which mitigates the benefits of variance-normalization as explained in [32]. A value of $\epsilon^* = 0.001$ was picked based on the performance of the WPE algorithm using oracle PSD.

## 2.3 DNN-based PSD estimation

The anechoic speech PSD estimate $\lambda^{(\mathrm{WPE})}$ is obtained at each time step, either by recursive smoothing of the reverberant periodogram [20] or with help of a DNN [21]. A block diagram of the DNN-WPE algorithm as proposed in [21] is given in Fig. 1, as the first stage up to $v^{(\mathrm{WPE})}$. In this approach, the input to the neural network is the magnitude of the reference channel $|x_0|$, taken here to be the first channel. We did not observe changes in the results by changing the reference channel or computing an average of the channels to obtain the DNN input, likely because the signal model itself considers a channel-agnostic PSD. The magnitude frame is then fed to a recurrent neural network MaskNet$_{\mathrm{WPE}}$, which outputs a real-valued mask $\mathcal{M}^{(\mathrm{WPE})}$. The PSD estimate is obtained by time-frequency masking:

$$\lambda_{t,f}^{(\mathrm{WPE})} = (\mathcal{M}_{t,f}^{(\mathrm{WPE})} \odot |x_{0,t,f}|)^2, \qquad (6)$$

where $\odot$ represents the Hadamard element product.

In [12, 21], the DNN is optimized with a mean-squared error (MSE) criterion on the masked output. In contrast, we proposed to use the $L^1$ loss:

$$\mathcal{L}_{\mathrm{DNN\text{-}WPE}} = \sum_{t,f} \left| \mathcal{M}_{t,f}^{(\mathrm{WPE})} \odot |x_{0,t,f}| - |v_{0,t,f}| \right|. \qquad (7)$$

This loss function indeed led to better results in our experiments [26]. This can be explained by the fact that the $L^1$ loss puts more weight on low-energy bins than high-energy bins in comparison to the MSE loss as it is more concave, which is a good fit for dereverberation.

## 2.4 End-to-end training procedure

### 2.4.1 End-to-end criterion and objectives

We argue that the mismatch between the DNN-optimization criterion (7) and the dereverberation task may limit the overall performance. However, using ASR as an end-to-end training criterion, as is done in [25], may not necessarily the best choice in order to optimize a dereverberation algorithm for hearing-aid users. The first reason is that the resulting scheme could not be adapted to specific user categories, although these benefit from different speech cues. Namely, HA listeners are shown to benefit from early reflections [9] where CI listeners do not significantly benefit from those, in particular in highly reverberant scenarios where early reflections degrade intelligibility [27]. The second reason is that by nature, the dereverberation scheme will provide the best representation possible for ASR, which may be not the optimal representation in terms of quality and intelligibility for a human listener.

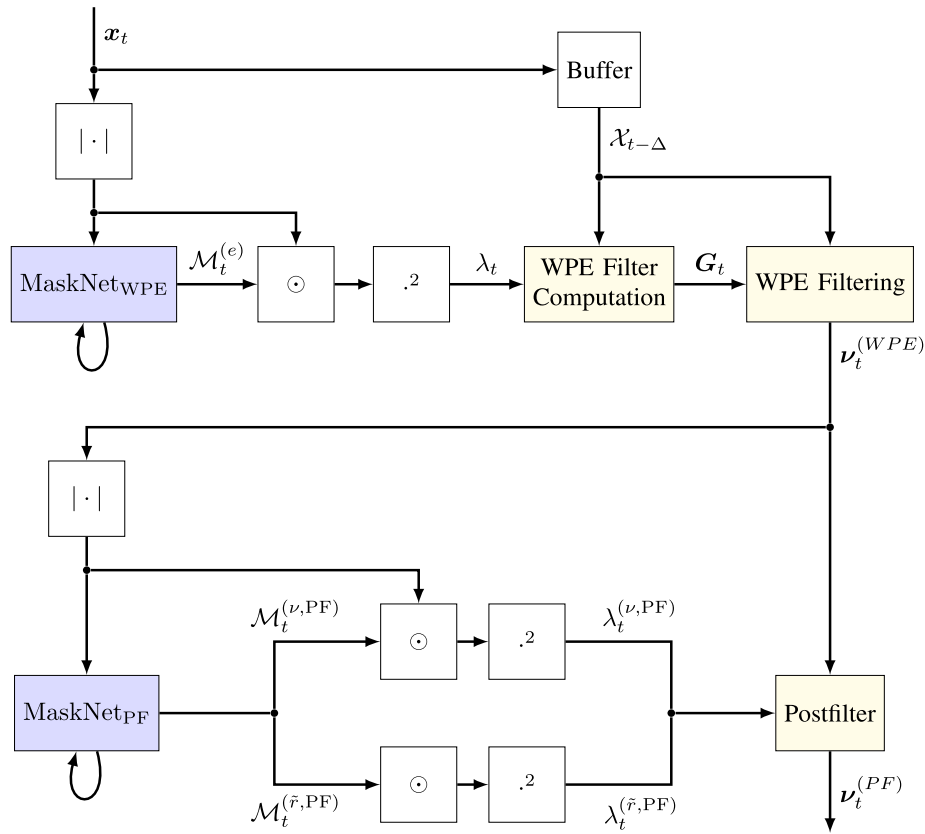

**Fig. 1** DNN-supported two-stage dereverberation. Blue blocks refer to trainable neural network layers. Yellow blocks represents adaptive statistical signal processing

We therefore proposed an end-to-end training procedure where the optimization criterion is placed in the time-frequency domain at the output of the DNN-WPE algorithm, thus including the back-end WPE into DNN optimization:

$$\mathcal{L}_{\text{E2E-WPE}} = \sum_{t,f} \left| |v_{t,f}^{(\text{WPE})}| - |v_{t,f}| \right|. \tag{8}$$

#### 2.4.2 End-to-end training procedure
An important practical aspect of this study focuses on handling the initialization period of the RLS-WPE algorithm. During this interval, the filter $G$ has not yet converged to a stable value, reducing dereverberation performance. Therefore, rather than relying on a hypothetical shortening of this period through implicit PSD optimization [25], we choose to exclude this initialization period from training. The DNN is thus optimized so that the algorithm works best in its stable regime. To do so, we first craft long reverberant utterances that we cut in segments of $L_i$ frames, where $L_i$ is the worst case initialization time plus some margin. We then design the training procedure so that the first segment is used only to initialize the WPE statistics $G$ and $R^{-1}$ and the DNN hidden states $h(\text{MaskNet}_{\text{WPE}})$. This enables to train the DNN weights on the next segments, during the stable regime. The data generation procedure is detailed again in subsection 4.

We showed in [26] that the best performance was obtained with the E2Ep-WPE approach, where the network MaskNet$_{\text{WPE}}$ is first pre-trained with (7) and fine-tuned with (8). If MaskNet$_{\text{WPE}}$ is only pre-trained, the algorithm is named DNN-WPE, and corresponds to [21] with a different training loss function.

The proposed end-to-end training procedure is summarized in Algorithm 1.

```
 1: Extract STFT of given sequence
 2: Segment sequence in N segments of size L_i
 3: for n ∈ {0 … N − 1} do
 4:    if n = 0 then          ▷ Initialization period
 5:        Initialize LSTM hidden state:
           h(MaskNet_WPE)_0^(0) = 0
 6:        Initialize WPE statistics:
           R^{-1}_0^{(0)} = I ;  G_0^{(0)} = 0
 7:        for t ∈ {0 … L_i − 1} do
 8:            Compute ν_t^(WPE)
 9:    if n > 0 then           ▷ After initialization
10:        Forward LSTM hidden state:
           h(MaskNet_WPE)_0^(n) = h(MaskNet_WPE)_{L_i−1}^{(n−1)}
11:        Forward WPE statistics:
           R^{-1}_0^{(n)} = R^{-1}_{L_i−1}^{(n−1)} ;  G_0^{(n)} = G_{L_i−1}^{(n−1)}
12:        for t ∈ {0 … L_i − 1} do
13:            Compute ν_t^(WPE)
14:        Backpropagate loss (8) through time on n
```

**Algorithm 1** End-to-End Training Procedure

## 3 Residual reverberation suppression
### 3.1 Signal model
As shown in Section 5, training the DNN-supported WPE stage in an end-to-end fashion helps suppressing large part of the reverberant signal immediately following the target range, that is, up to $L_m$, which we refer to as the *moderate reverberation* range.

We thus refine the reverberant signal model as (1):

$$x_t = v_t + m_t + \phi_t + u_t, \qquad (9)$$

where the undesired reverberant signal in (1) (corresponding to $r$ and $e + r$ in the HA and CI case respectively) is split in the *moderate* reverberant signal $m$ and the *final* reverberant signal $\phi$, defined as:

$$m_t = \sum_{\tau=\Delta}^{\Delta+L_m-1} h_\tau s_{t-\tau}, \qquad (10)$$

$$\phi_t = \sum_{\tau=\Delta+L_m}^{N} h_\tau s_{t-\tau}. \qquad (11)$$

The resulting WPE estimate thus contains the target $v$, a target estimation error $\tilde{v}$, a residue $\tilde{m}$ from this moderate reverberation and a residue stemming from the final reverberation $\tilde{\phi}$ (again disregarding the error term $u$ in noiseless scenarios):

$$\begin{aligned}v_t^{(\mathrm{WPE})} &= x_t - G_t^H \mathcal{X}_{t-\Delta} \\ &= v_t + \underbrace{\tilde{v}_t + \tilde{m}_t + \tilde{\phi}_t}_{\tilde{r}_t}.\end{aligned} \qquad (12)$$

The target estimation error $\tilde{v}$ is the target component which was degraded by the algorithm. As described in [32] for the original WPE algorithm, parts of the early reflections may be destroyed because of the inner short-time speech correlations. Under some mild assumptions, the direct path is however fully preserved if the prediction delay $\Delta$ is sufficiently large (i.e., larger than the inner speech correlation time). The target estimation error is therefore likely to be larger when using WPE-based algorithms in the HA scenario—containing more early reflections—than in the CI scenario.

### 3.2 Postfiltering scheme
We aim at suppressing the two residues $\tilde{m}$ and, more particularly, $\tilde{\phi}$. Indeed, $\tilde{\phi}$ is generally of higher magnitude than $\tilde{m}$, as we will show in the experiments that a large amount of moderate reverberation can be canceled by efficient WPE-based dereverberation. Additionally, $\tilde{\phi}$ is the more perceptually disturbing of the two residues for the following reasons.

On the one hand, $\tilde{\phi}$ can be considered as speech-like noise which is very poorly correlated to the target signal in comparison to $\tilde{m}$. On the other hand, as WPE cancels most of the so-called moderate reverberation, there is no preceding energy anymore to mask the late reverberation. The final reverberation residue is then clearly audible.

We thus add a post-filtering enhancement stage after the linear WPE filtering stage, which consists of a single-channel Wiener filter, the phase being left unchanged. This Wiener filter uses estimates of the target PSD $\lambda^{(v,\mathrm{PF})}$ and interference PSD $\lambda^{(\tilde{r},\mathrm{PF})}$, which can be obtained with classical techniques as decision-directed signal-to-noise ratio (SNR) estimation [33], cepstral smoothing [6, 34], or from a neural network [21, 35].

The resulting estimate is then given for each channel $d$ separately by the celebrated Wiener filter, using the WPE output:

$$v_{d,t}^{(\mathrm{PF})} = \frac{\lambda_{d,t}^{(v,\mathrm{PF})}}{\lambda_{d,t}^{(v,\mathrm{PF})} + \lambda_{d,t}^{(\tilde{r},\mathrm{PF})}} v_{d,t}^{(\mathrm{WPE})} \qquad (13)$$

### 3.3 DNN-based PSD estimation
We use a DNN-based masking approach to obtain the target and residual reverberation PSDs, similar to what is used to estimate the target speech PSD for WPE filtering (see (6)). At each time step, a frame of the WPE output's magnitude taken from the reference channel $|v_0^{(\mathrm{WPE})}|$ is fed to a recurrent neural network MaskNet$_{\mathrm{PF}}$, which outputs both a target and interference mask. The PSD estimate $\lambda^{(\eta)}$ is then obtained for each channel $d$ through time-frequency masking for each signal $\eta \in \{v, \tilde{r}\}$:

$$\lambda_{d,t,f}^{(\eta)} = (\mathcal{M}_{t,f}^{(\eta)} \odot |v_{d,t,f}^{(\mathrm{WPE})}|)^2. \quad (14)$$

We apply the same reference-channel mask for all channels using only one instance of the DNN, which saves some computational power and enables us to leave the interaural level differences unchanged. Also, the interaural phase differences are well estimated by WPE linear filtering and are not modified by the post-filtering scheme (see (13)). Therefore the target binaural cues are well preserved, which is important for hearing devices.

A block diagram of the complete two-stage algorithm is provided in Fig. 1.

### 3.4 Training procedure

We trained the post-filter DNN MaskNet$_{\mathrm{PF}}$ with a similar mask-based objective as MaskNet$_{\mathrm{WPE}}$:

$$\mathcal{L}_{\mathrm{DNN\text{-}PF}} = \sum_{t,f} \left| \mathcal{M}_{t,f}^{(v)} \odot |v_{0,t,f}^{(\mathrm{WPE})}| - |v_{0,t,f}| \right| \\ + \sum_{t,f} \left| \mathcal{M}_{t,f}^{(\tilde{r})} \odot |v_{0,t,f}^{(\mathrm{WPE})}| - |\tilde{r}_{0,t,f}| \right|, \quad (15)$$

where $\tilde{r}_0$ is the undesired signal defined in (12) taken at the reference channel. We report results for two approaches. First is DNN-WPE+DNN-PF, where the network MaskNet$_{\mathrm{WPE}}$ is pre-trained with (7), then frozen for the pre-training of MaskNet$_{\mathrm{PF}}$ with (15). Second is E2Ep-WPE+DNN-PF, where the network MaskNet$_{\mathrm{WPE}}$ is pre-trained with (7) and fine-tuned with (8), then frozen for the pre-training of MaskNet$_{\mathrm{PF}}$ with (15).

A table making the present algorithms correspond to their characteristics and acronyms is given in Table 1.

## 4 Experimental Setup
### 4.1 Dataset generation

We use clean speech material from the WSJ0 dataset [36], using the usual split of 101, 10, and 8 speakers for training, validation, and testing respectively. For each split independently, we concatenate utterances belonging to the same speaker, and construct sequences of approximately 20 s. The initialization time of WPE can go up to to 2 s in the worst case when using a forgetting factor of $\alpha = 0.99$. For end-to-end training, we do not want to learn during that period (cf Section 2.4). Therefore, we cut these long sequences in segments of $L_i = 4$ s and use the first segment only for initialization, thus not backpropagating the loss on it (cf Algorithm 1). We choose $L_i$ to fill both requirements of (i) being larger than the worst case initialization time of WPE and (ii) providing a sufficient receptive field for training with LSTMs. Since the first segment is never used for optimization, permutations of the original utterances are used to create several versions of each sequence, so that we still use all speech data available for training the DNNs.

**Table 1** List of acronyms for strategies estimating the PSD used in the linear filtering and non-linear post-filtering stages

| Algorithm | $\lambda^{(\mathrm{WPE})}$ | $\lambda^{(v,\mathrm{PF})}, \lambda^{(\tilde{r},\mathrm{PF})}$ |
| --- | --- | --- |
| RLS-WPE [20] | Reverberant | ✗ |
| O-PSD-WPE | Oracle | ✗ |
| DNN-PF | ✗ | $\mathcal{L}_{\mathrm{DNN-PF}}$ |
| DNN-WPE | $\mathcal{L}_{\mathrm{DNN-WPE}}$ | ✗ |
| E2Ep-WPE | $\mathcal{L}_{\mathrm{DNN-WPE}} \to \mathcal{L}_{\mathrm{E2E-WPE}}$ | ✗ |
| DNN-WPE+DNN-PF | $\mathcal{L}_{\mathrm{DNN-WPE}}$ | $\mathcal{L}_{\mathrm{DNN-PF}}$ |
| E2Ep-WPE+DNN-PF | $\mathcal{L}_{\mathrm{DNN-WPE}} \to \mathcal{L}_{\mathrm{E2E-WPE}}$ | $\mathcal{L}_{\mathrm{DNN-PF}}$ |

These sequences are convolved with 2-channel RIRs generated with the RAZR engine [37] and randomly picked. Each RIR is generated by uniformly sampling room acoustics parameters as in [30] and a $T_{60}$ reverberation time between 0.4 and 1.0 s. Head-Related Transfer Function based auralization is performed in the RAZR engine, using a KEMAR dummy head response from the MMHR-HRTF database [38].

As specified earlier, the target data for the HA case should represent the direct path and the early reflections as normal hearing and hearing-aided listeners benefit from early reflections [9]. Therefore, we convolve the dry utterance with the beginning of the RIR, up to a separation time often found in the dereverberation literature [1, 9, 39]. We empirically set the separation time to 40 ms instead of the usual 50 ms, as we obtained better instrumental results when comparing the resulting target data to WPE estimates using the oracle PSD.

In the CI scenario, the target data data should theoretically contain the direct path only [27]. However, directly estimating the direct path from reverberant speech often provides poor instrumental results given the low input SNR. Note also that the first WPE stage uses a prediction delay $\Delta$ supposed to protect the inner speech correlations, whose range is usually estimated to $\sim 10$ ms. The minimal $\Delta$ that fills this requirement is $\Delta = 2$ STFT frames with the hyperparameters described below, that is, 16 ms. Therefore, we propose to match the target data with the best possible WPE estimate, by convolving the dry utterance with the first 16 ms of the RIR. This also contributes to decreasing the difficulty of the estimation task, which helps obtain reasonable estimates with the proposed algorithm. We further noticed that with this setting, very few early reflections could be heard in the target.

The original mean input direct-to-reverberant ratio (DRR) between the dry signal and reverberant mixture is $-6.0$dB and the mean microphone-to-speaker distance used was estimated to 4.2m. The resulting mean input signal-to-noise ratio (SNR) between the generated target and the reverberant mixture is 0.9dB for the HA scenario, and $-1.4$dB for the CI scenario.

Finally, independent and identically distributed Gaussian noise is added to each channel with an input SNR uniformly sampled in [15, 25] dB to simulate sensor noise. Ultimately, the training, validation and testing sets contain around 55, 16 and 3 h of speech sampled at 16 kHz.

### 4.2 Hyperparameter settings

The STFT uses a square-rooted Hann window of 32 ms and a 75 % overlap. For training, segments of $L_i = 4$ s are constructed from each sequence (see Section 4.1). All approaches are trained using the Adam optimizer with a learning rate of $10^{-4}$ and a batch size of 128. Training is stopped if a maximum of 500 epochs is reached or if early stopping is detected, in case the validation loss has not decreased in 20 consecutive epochs.

The WPE filter length is set to $K = 10$ STFT frames (i.e., 80 ms), the number of channels to $D = 2$, the WPE adaptation factor to $\alpha = 0.99$, and the delays to $\Delta_{HA} = 5$ frames (i.e., 40 ms) for the HA scenario and $\Delta_{CI} = 2$ (i.e., 16 ms) frames for the CI scenario. The delay values are picked to match the amount of early reflections contained in the respective target, and they experimentally provide optimal evaluation metrics when comparing the corresponding target to the output of WPE when using the oracle PSD (see Section 4.1).

The DNN used in [21] is composed of a single long-short term memory (LSTM) layer with 512 units followed by two linear layers with rectified linear activations (ReLU) and a linear output layer with sigmoid activation. We remove the two ReLU-activated layers in our experiments, which did not significantly degrade the dereverberation performance, while reducing the number of trainable parameters by 75 %, therefore ending with 1.6M parameters. We use the same architecture for MaskNet$_{\text{WPE}}$ and MaskNet$_{\text{PF}}$. We choose to use LSTMs rather than recent convolutional network- or transformer-based architectures to develop a frugal algorithm for hearing devices with limited computing resources. Indeed, LSTMs require much fewer operations per second than the mentioned alternatives, given that they process only one input frame and perform sequence-modeling using their internal memory state.

### 4.3 Evaluation metrics

We evaluate all approaches on the described test sets corresponding to the HA and CI scenarios.

Following the definition of the early-to-late reverberation ratio (ELR) [10, 40], we introduce two new instrumental measures: the *early-to-moderate reverberation ratio* (EMR) and *early-to-final reverberation ratio* (EFR). Estimated RIR coefficients $\{\hat{H}\}_{d,\tau,f}$ of order $0 \leq \tau \leq P-1$ are computed for each channel $d$ and frequency bin $f$ separately, in order to minimize a minimum mean square error regression objective in the time-frequency domain between a reverberant utterance $Y$ and the corresponding dry utterance $S$ filtered by $H$ [13]:

$$\{\hat{H}_{d,\tau,f}\}_\tau = \arg\min_H \sum_{t=0}^{T-1} ||Y_{d,t,f} - \sum_{\tau=0}^{P-1} H_{d,\tau,f} S_{t-\tau-\delta^*,f}||_2^2, \quad (16)$$

with $\delta^*$ being the oracle propagation delay obtained by looking for the direct path in the true RIR. This delay is used so as not to try and estimate RIR coefficients preceding the propagation delay which are supposed to be zero, therefore reducing the estimation error. The estimation error is further reduced by choosing the order $P$ to match the $T_{30}$ of the true RIR rather than the $T_{60}$, as the estimation error floor was found to be close to $-30$dB.

The channel-wise RIRs are then stacked and the target, moderate and final reverberation components are estimated as:

$$\widehat{v}_{t,f} = \sum_{\tau=0}^{\tilde{\Delta}-1} \hat{H}_{\tau,f} S_{t-\tau-\delta^*,f}, \quad (17)$$

$$\widehat{m}_{t,f} = \sum_{\tau=\tilde{\Delta}}^{\tilde{\Delta}+L_m-1} \hat{H}_{\tau,f} S_{t-\tau-\delta^*,f}, \quad (18)$$

$$\widehat{\phi}_{t,f} = \sum_{\tau=\tilde{\Delta}+L_m}^{P-1} \hat{H}_{\tau,f} S_{t-\tau-\delta^*,f}. \quad (19)$$

We set $\tilde{\Delta} = 5$ (i.e., 40ms) in the hearing-aided case and $\tilde{\Delta} = 2$ (i.e., 16ms) in the cochlear-implanted scenario as explained in the target specifications in the section above. We set the moderate range length to $L_m = K = 10$ (i.e., 80ms).

The ELR, EMR and EFR are then defined as:

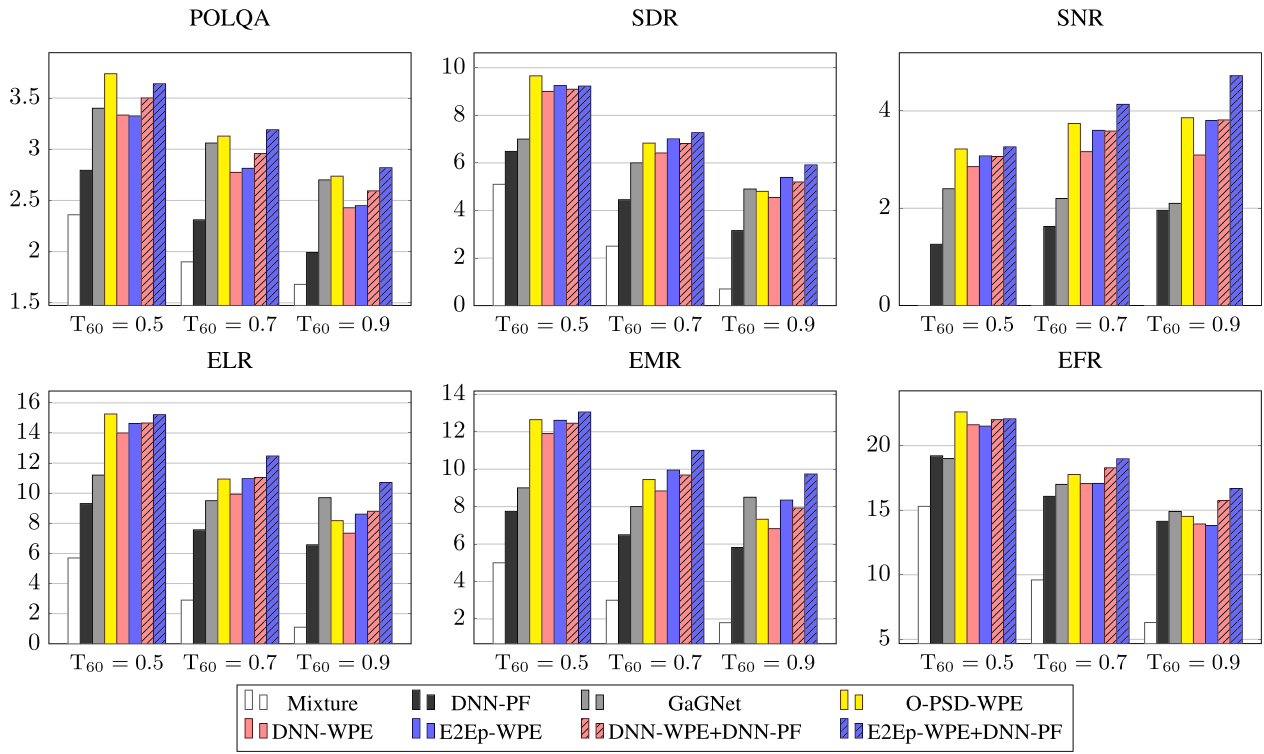

**Fig. 2** Scores on unprocessed and processed signals for hearing-aided scenario. All metrics except POLQA are in dB. $T_{60}$ times indicated in s. [$v = d + e$; $\Delta = 5$]

$$\mathrm{ELR} = 10 \log_{10} \left( ||\hat{v}||^2 / ||\hat{m} + \hat{\phi}||^2 \right), \quad (20)$$

$$\mathrm{EMR} = 10 \log_{10} \left( ||\hat{v}||^2 / ||\hat{m}||^2 \right), \quad (21)$$

$$\mathrm{EFR} = 10 \log_{10} \left( ||\hat{v}||^2 / ||\hat{\phi}||^2 \right). \quad (22)$$

We complete the evaluation benchmark with Perceptual Objective Listening Quality Analysis (POLQA)[1], signal-to-distortion ratio (SDR), and signal-to-noise ratio [41].

## 5 Experimental results and discussion
### 5.1 Compared algorithms
We apply the different strategies mentioned in Sections 2 and 3 and compare their results in Figs. 2 and 3 for the HA and CI scenarios of our simulated dataset respectively.

Spectrograms are also plotted in Fig. 4. We add to the already proposed approaches (mentioned in italics):

- *O-PSD-WPE*: RLS-WPE using the oracle target PSD
- *DNN-PF*: The output of the network $\mathrm{MaskNet}_{\mathrm{WPE}}$ is directly used for single-channel Wiener non-linear filtering, eluding the WPE linear filter step
- *GaGNet* [42]: A recent CNN-based network for hybrid magnitude and complex domain enhancement. GaGNet is the successor of [43] which was ranked first in the real-time enhancement track of the DNS-2021 challenge [44]. We used the open source available implementation[2] but adapted the number of frequency bins to be 257 as in our implementation

Some listening examples and spectrograms are available on our dedicated webpage[3]. We also include there a video recording of our proposed E2Ep-WPE+DNN-PF (HA) algorithm performing in real time in both static and moving speaker scenarios. The algorithm performs with a total latency of 40 ms determined by the 32 ms *algorithmic latency* due to the STFT synthesis window length and the 8ms *processing time* which is contained within a STFT hop. We show that for reasonable speaker movements, the algorithm yields high performance also in the dynamic setting.

---

[1] Wideband MOS score, following standard ITU-T P.863. The authors would like to thank Rohde & Schwarz SwissQual AG for their support with POLQA.

[2] https://github.com/Andong-Li-speech/GaGNet

[3] https://uhh.de/inf-sp-twostagederev

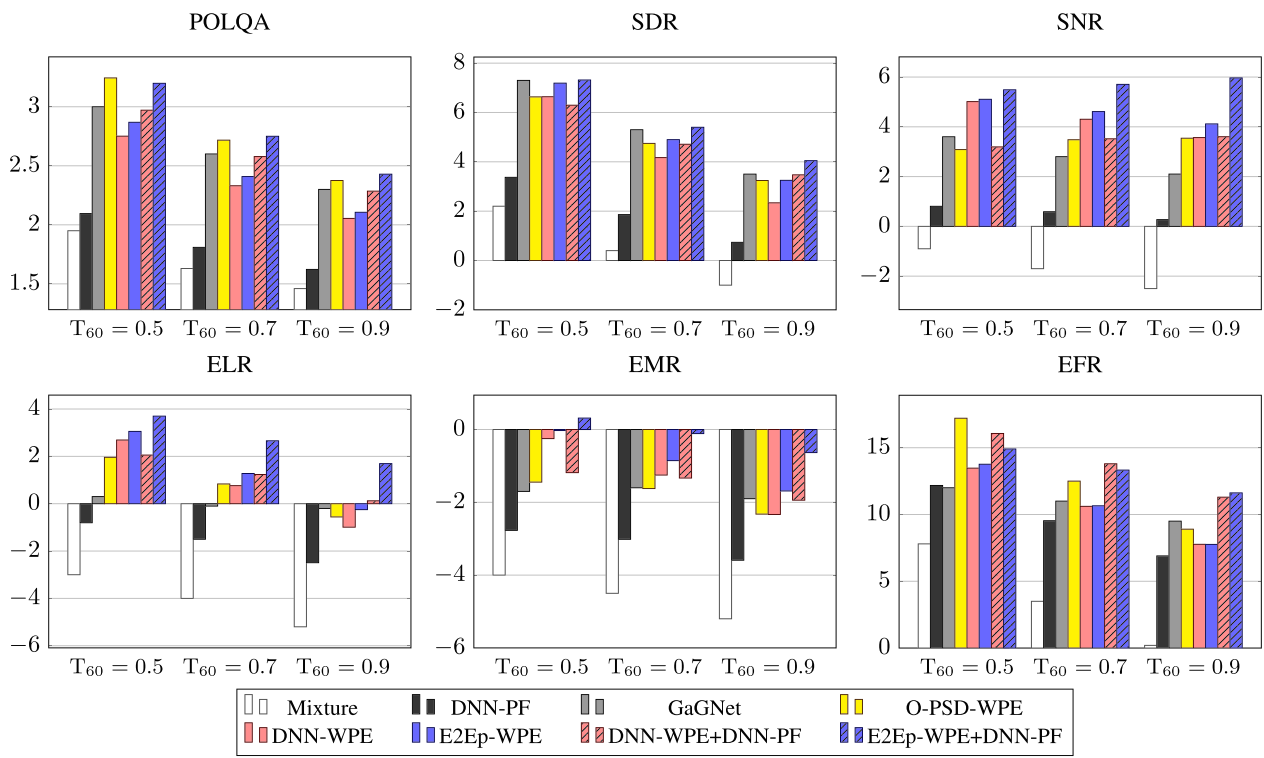

**Fig. 3** Scores on unprocessed and processed signals for cochlear-implanted scenario. All metrics except POLQA are in dB. $T_{60}$ times indicated in s. [$v = d; \Delta = 2$]

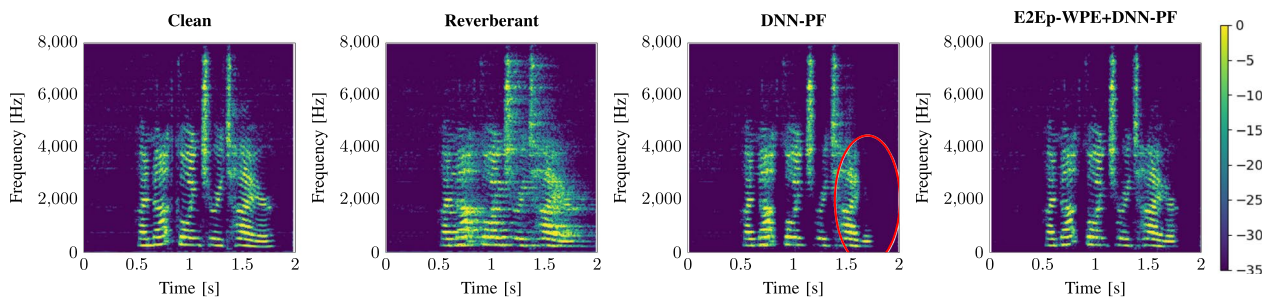

**Fig. 4** Log-energy spectrograms of clean, reverberant, and processed utterances. $T_{60} = 0.68$s. HA scenario [$v = d + e; \Delta = \tilde{\Delta} = 5$] Heavy speech distortions can be observed in the DNN-PF output, as highlighted in the red ellipse

### 5.2 Moderate reverberation suppression

We first validate the method used for deriving the ELR, EMR and EFR metrics, described in 4.3. We plot the log-energies of the true RIR, the RIR estimated with (16) and the transfer function of the concatenation of the room with the O-PSD-WPE algorithm on Fig. 5. We observe that in the chosen $T_{30}$ range, the true and estimated RIRs match almost perfectly, showing the validity of this MMSE-based estimation for linear transfer function estimation in this range. We also observe a strong dereverberation performance of the O-PSD-WPE algorithm in the filter range as well as shortly after this range, which is the effect of recursive averaging.

The ELR metric in Figs. 2 and 3 indicates a superior dereverberation performance of E2Ep-WPE in comparison to DNN-WPE, i.e., when the DNN MaskNet$_{\text{WPE}}$ is fine-tuned end-to-end. The high EMR difference indicates that the moderate reverberation in the range

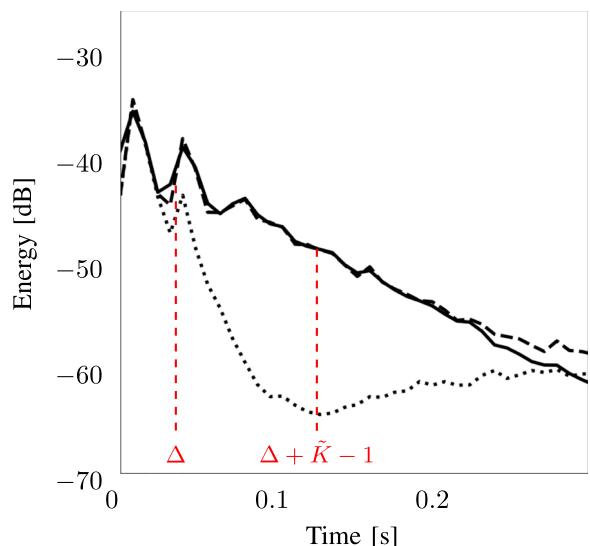

**Fig. 5** Comparison of the true RIR (full line) vs. the estimated RIR (dashed line). Estimated linearized transfer function of the system which applies O-PSD-WPE on reverberant speech is shown as a dotted line. We observe strong dereverberation in the given filter range $[\Delta, \Delta + \tilde{K} - 1]$ and shortly afterwards because of recursive averaging. Only the $T_{30}$ range is displayed as it is the valid estimation range for the estimated RIR. $T_{60} = 0.8$s. HA scenario $[\mathbf{v} + \mathbf{d} + \mathbf{e}; \Delta = \tilde{\Delta} = 5]$

$[\tilde{\Delta}, \tilde{\Delta} + L_m - 1]$ is particularly well suppressed. As already mentioned in [26], this stems from the better dereverberation performance in the range which is available to the WPE linear filter, through end-to-end optimization of the neural network MaskNet$_{\text{WPE}}$.

### 5.3 Residual reverberation suppression

As displayed in Figs. 2 and 3, using a DNN-assisted post-filtering stage highly improves the dereverberation performance on the basis of WPE linear filtering, and yields much superior POLQA scores. The high EFR improvement indicates that post-filtering mostly focuses on removing the final reverberation, i.e., after the range accessible to WPE filtering. In particular, the E2Ep-WPE+DNN-PF approach which uses a pretrained network for post-filtering on top of end-to-end trained WPE filtering outperforms all other approaches on all metrics. In comparison, using only the post-filter without WPE filtering introduces a lot of speech distortion, as shown in Fig. 4. Similarly, the DNN-WPE+DNN-PF performance indicates that using the post-filtering stage on the output of the DNN-WPE algorithm—without fine-tuning MaskNet$_{\text{WPE}}$ with our end-to-end procedure—yields poorer results (final POLQA is 0.2 lower and SNR is 1dB lower than E2Ep-WPE+DNN-PF). This shows that removing the moderate reverberation with WPE linear filtering is an essential step before using a speech enhancement scheme like our post-filter. Since E2Ep-WPE efficiently removes the moderate reverberation, as measured by EMR, it provides a particularly good ground for enhancement-like post-filtering, since only the reverberation tail remains and provides the best EFR and POLQA performance.

### 5.4 Reverberation times

For a given scenario, the dereverberation task becomes increasingly difficult as the $T_{60}$ time grows longer. We observe for example that using the oracle PSD for WPE performs well only for low $T_{60}$ reverberation times because of the limited filter length, and the performance gap between this approach and the proposed two-stage approach increases with the $T_{60}$ reverberation time.

Furthermore, we notice an increasing gap in SNR and EFR between DNN-WPE+DNN-PF and E2Ep-WPE+DNN-PF as the $T_{60}$ grows larger, which seems to indicate that our best performing approach E2Ep-WPE+DNN-PF is more robust to challenging reverberation conditions.

### 5.5 Hearing device users categories specialization

Similar trends in performances are observed for the hearing-aided and cochlear-implanted scenarios.

Dereverberation is a more complicated task in the CI scenario as compared to the HA scenario, as the input ELR and SDR scores are lower. Yet, the POLQA and SDR score improvements stay relatively consistent across both scenarios, highlighting the robustness of our approach. However, the EMR improvements seem larger in the HA scenario than in the CI scenario. Indeed, it is more arduous in the latter scenario to remove the beginning of what is considered to be the reverberant tail, as it includes parts of the early reflections, which are complicated to attenuate without degrading the direct path. This also accounts for the smaller EMR improvement of E2Ep-WPE over DNN-WPE, as compared to the HA scenario. Furthermore, the SNR improvements are larger in the CI scenario than in the HA scenario, especially those brought by the proposed E2Ep-WPE+DNN-PF approach, which shows that the post-filtering stage is in this case able to remove a lot of the residual reverberation.

### 5.6 Computational requirements

We estimate the number of MAC operations per second of the models using the `python-papi` Python package which provides CPU counters for single- and double-point precision operations. We end up with an estimate of 0.13 GMAC·s$^{-1}$ for our proposed E2Ep-WPE+DNN-PF algorithm running at 16 kHz. With the same estimation method, the implemented GaGNet uses 0.81 GMAC·s$^{-1}$. Also with regard to memory, our method has a lower budget as GaGNet has 11.8M trainable parameters while our approach has 3.2M parameters.

Our method therefore outperforms GaGNet on the proposed dataset with a significantly smaller computational load, without special fine-tuning of the hyperparameters nor optimization of the architectures used.

## 6 Conclusions

We have proposed a lightweight two-stage DNN-assisted algorithm for frame-online adaptive multi-channel dereverberation on hearing devices. The first stage consists of multi-frame, multi-channel linear filtering with help of a DNN estimating the target speech PSD, optimized end-to-end. This first stage was shown to focus on accurately removing moderate reverberation up to the given filter range, in our case, 120 ms. The second stage performs channel-wise, single-frame non-linear spectral enhancement with help of a DNN estimating the target and interference PSDs. This second stage is able to efficiently remove residual late reverberation left off by the first stage.

Our model-based approach allows to tailor the two-stage algorithm toward different classes of hearing-impaired listeners, namely hearing-impaired listeners benefiting from early reflections on the one hand, and cochlear-implanted users on the other hand benefiting from the direct path only.

Instrumental metrics like the early-to-late reverberation ratio and its variants confirm the listening-based experiments showing the complementary aspect of the two proposed stages.

The proposed approach outperforms a state-of-the-art DNN-based enhancement scheme on the proposed dataset, using a significantly smaller time and memory footprint.

### Abbreviations
| | |
|---|---|
| DNN | Deep neural network |
| WPE | Weighted prediction error |
| PSD | Power spectral density |
| SNR | Signal-to-noise ratio |
| RLS | Recursive least squares |
| ASR | Automatic speech recognition |
| HA | Hearing aid |
| CI | Cochlear implant |
| DRR | Direct-to-reverberant ratio |

**Acknowledgements**
Not applicable.

**Authors' contributions**
All authors listed have contributed significantly to this work. The authors read and approved the final manuscript.

**Funding**
Open Access funding enabled and organized by Projekt DEAL. This work has been funded by the Federal Ministry for Economic Affairs and Climate Action, project 01MK20012S, AP380. The authors are responsible for the content of this paper.

**Availability of data and materials**
The data that support the findings of this study are available from the Linguistic Data Consortium but restrictions apply to the availability of these data, which were used under license for the current study, and so are not publicly available. Data are however available from the authors upon reasonable request and with permission of Linguistic Data Consortium.

## Declarations

**Competing interests**
The authors declare that they have no competing interests.

Received: 24 November 2022   Accepted: 11 April 2023
Published online: 01 May 2023

## Publisher's Note

Springer Nature remains neutral with regard to jurisdictional claims in published maps and institutional affiliations.